\documentclass[a4paper]{jpconf}
\usepackage{graphicx,amsmath,amssymb}

\newcommand{\AIP}{{\it AIP Conf. Proc.} }
\newcommand{\CP}{{\it Chin. Phys.} }
\newcommand{\EPJ}{{\it Eur. Phys. J.} }

\newcommand{\JPCS}{{\it J. Phys.: Conf. Ser.} }

\newcommand{\PRPT}{{\it Phys. Rep.} }
\newcommand{\PREP}{{\it Preprint\/} }

\begin{document}

\title{Pentaquark photoproduction}

\author{C\'esar Fern\'andez-Ram\'{\i}rez}
\address{Instituto de Ciencias Nucleares, 
Universidad Nacional Aut\'onoma de M\'exico,
Ciudad de M\'exico 04510, Mexico}
\ead{cesar.fernandez@nucleares.unam.mx}

\author{Astrid N.~Hiller Bin}
\address{Departamento de F\'{\i}sica Te\'orica and IFIC, 
Centro Mixto Universidad de Valencia-CSIC, 
Institutos de Investigaci\'on de Paterna, E-46071 Valencia, Spain}

\author{Alessandro Pilloni}
\address{Theory Center, 
Thomas Jefferson National Accelerator Facility, 
Newport News, VA 23606, USA}

\begin{abstract}
We present results and suggestions on how to confirm the existence and resonant
nature of the $Pc(4450)$ detected at LHCb through photoproduction experiments. 
We find that this narrow structure might have escaped detection in past experiments and use those to give a constraint 
for the upper limit of the branching ratio/coupling to the $J/\psi\: p$ channel.
\end{abstract}

\section{Introduction}
One of the most remarkable properties of Quantum Chromodynamics (QCD) 
is the color confinement phenomenon that
makes hadrons to be built out of color singlets of quarks.
The simplest way to build a color neutral baryon (half-integer spin) state is out of three quarks ($qqq$).
However, QCD tells us that there are other ways  to build color neutral baryons, for example: 
hybrid  baryons with gluonic components ($qqqg$, $qqqgg$, \dots),
pentaquarks ($qqqq\bar{q}$), heptaquarks ($qqqqq\bar{q}\bar{q}$) and so on. 
Finding complete multiplets of baryons beyond the $qqq$ picture is one of the main goals of hadron spectroscopy
and will provide insight on the dynamics of QCD in the hadronic energy regime.
The recent discovery of two structures compatible with pentaquark states in the $\Lambda_b^0 \to J/\psi K^- p$
decay by LHCb~\cite{LHCbpentaquark} 
together with the observation of a compatible structure in $\Lambda_b^0 \to J/\psi\,\pi^- p$~\cite{LHCb2},
has triggered a lot of research activities aimed to explain them.
Each signal has been associated to a pentaquark candidate, $P_c(4380)$ and $P_c(4450)$, with opposite parity assignments, 
and one state having spin $3/2$ and the other $5/2$~\cite{LHCbpentaquark,PDG2016}.
The $P_c(4380)$ signal is less apparent and the properties of this state (mass and width) 
could depend on the amplitude analysis performed, 
especially for the $K^-p$ background~\cite{PDG2016,Fernandez-Ramirez:2015tfa}.
However, the $P_c(4450)$ signal is very narrow and its mass and width 
are not expected to change by refinements in the amplitude analysis.
Hence, its study has been the focus of many research groups.
Several explanations for this signal have been proposed: a kinematical effect~\cite{kinematical}, 
a molecular state~\cite{molecule} 
and a compact QCD (pentaquark) state~\cite{pentaquark}.

One of the best approaches to gain insight on the nature of $P_c(4450)$ consists 
on detecting the state in a different reaction.
In~\cite{Wang:2015jsa} it was proposed to use $J/\psi\, p$ photoproduction. 
From the LHCb $\Lambda_b^0$ decay data
we know that the $P_c(4450)$ state couples strongly to the $J/\psi\, p$, and 
the photon energy can be tuned to excite this particular state.
This experiment has just been approved to run at Hall C in Jefferson Lab~\cite{PcJLab}  
and results are expected in the next years.
Hence, as a first step it is necessary to develop the amplitudes that will be needed to provide realistic 
predictions for the planning of the experiment and later to analyze the experimental data.
In the next section we sketch the main features of the $J/\psi$ 
photoproduction model that we developed with these purposes in mind. 
The detailed description of the model and the corresponding results
can be found in~\cite{ANHB2016a}.

\section{Pentaquark photoproduction model}
The differential cross section for the $\gamma p \to J/\psi\, p$ reaction can be written
\begin{equation}
\label{Edsigdcos}
\frac{d\sigma}{d \cos\theta}= \frac{ 4\pi\alpha}{32 \pi  s} \frac{p_f}{p_i} \frac{1}{4} 
\sum_{\lambda_\gamma,\lambda_p, \lambda_{\psi}, \lambda_{p'}}
\left|   \langle \lambda_{\psi}\lambda_{p^\prime} \left|T\right|\lambda_\gamma \lambda_p \rangle   \right|^2 ,
\end{equation}
where $p_i$ is the incoming momentum, $p_f$ is the outgoing momentum, $\theta$ is the scattering angle,
$s$ is the standard Mandelstam variable, $\alpha$ is the electromagnetic fine structure constant, and
$\lambda_{\psi}$, $\lambda_{p^\prime}$, $\lambda_\gamma$ and  $\lambda_p$ are the
$J/\psi$, outgoing proton, photon and incoming proton helicities respectively.
All the kinematical variables are defined in the center of mass reference system.

The amplitude $\langle \lambda_{\psi}\lambda_{p^\prime} \left|T\right|\lambda_\gamma \lambda_p \rangle$
is constructed as the addition of two pieces:
(i) the pentaquark resonance as a Breit--Wigner amplitude
\begin{align}
\langle \lambda_{\psi}\lambda_{p^\prime} | T_r | \lambda_\gamma \lambda_p \rangle =
\frac{\langle \lambda_{\psi} \lambda_{p^\prime} | T_\text{dec} | \lambda_r  \rangle
\langle \lambda_r | T^\dagger_\text{em} | \lambda_\gamma \lambda_p \rangle}
{M_r^2-W^2-\mathrm{i}\Gamma_rM_r},\label{EBWamp}
\end{align}
where $M_r$, $\Gamma_r$ and $\lambda_r$ are the $P_c(4450)$ mass, width and helicity respectively, 
$W=\sqrt{s}$ is the invariant mass,
$T_\text{dec}$ stands for the strong vertex $p J/\psi P_c$, 
and $T_\text{em}$ stands for the electromagnetic vertex $\gamma p P_c$;
and (ii) a nonresonant background described through 
an affective Pomeron exchange that is constrained by existing high-energy data~\cite{data}
\begin{equation}
\langle \lambda_\psi\lambda_{p^\prime}|T_{P} |\lambda_\gamma \lambda_p\rangle = 
iA~\left(\frac{s-s_t}{s_0}\right)^{\alpha(t)}~e^{b_0(t-t_\text{min})}
\delta_{\lambda_p\lambda_{p'}}\delta_{\lambda_{\psi}\lambda_\gamma},
\label{EqPom}
\end{equation}
where $t$ is the standard Mandelstam variable, $t_\text{min}$ is the minimum $t$ for a given $s$, 
$s_0 = 1~\text{GeV}^2$ is the hadron physics scale, 
$\alpha(t)=\alpha_0+\alpha' t$ is the Pomeron Regge trajectory whose parameters $\alpha_0$ and $\alpha'$ are fitted to the data, 
and $A$, $b_0$ and $t_0$  are also fitted to the data.
The electromagnetic vertex can be related to the helicity amplitudes $A_{1/2}$ and $A_{3/2}$ through
\begin{equation}
\label{a12a32sec}
\langle \lambda_{\gamma}\lambda_{p} \left| T_\text{em} \right| \lambda_{r} \rangle  
=\frac{W}{M_{r}}\sqrt{\frac{8M_{N}M_{r} \bar p_{i}}{4\pi\alpha}}
\sqrt{\frac{\bar p_i}{p_i}}A_{1/2,3/2}\: ,
\end{equation}
where $M_N$ is the nucleon mass and $\bar p_{i}$ is the momentum $p_i$
evaluated at the resonance mass $M_r$.
We do not know the helicity amplitudes of the pentaquark,
hence, we need to make assumptions to describe this piece of the model.
We assume the electromagnetic and strong vertices to be related by vector meson dominance
\begin{equation}
\langle \lambda_{\gamma}\lambda_p|T_\text{em} | \lambda_r \rangle = 
\frac{\sqrt{4\pi \alpha} f_\psi}{M_\psi} 
\langle \lambda_{\psi}=\lambda_\gamma ,\lambda_p |T_\text{dec} | \lambda_r \rangle, \label{vmd}
\end{equation}
where $M_\psi$ and $f_\psi=280~\text{MeV}$ are the $J/\psi$ mass and decay constant respectively, and
\begin{align}
\langle \lambda_{\psi}\lambda_{p'} |T_\text{dec} | \lambda_r \rangle=
g_{\lambda_{\psi}\lambda_{p'}} d_{\lambda_r,\lambda_{\psi}-\lambda_{p'}}^{J_r}(\cos\theta), 
\end{align}
where $d_{\lambda_r,\lambda_{\psi}-\lambda_{p'}}^{J_r}(\cos\theta)$ are the Wigner $d$-functions
with $J_r$ the resonance spin, and $g_{\lambda_\psi, \lambda_{p'}}$ are the helicity couplings.
We assume the three helicity couplings to be equal,
$g=g_{\lambda_\psi, \lambda_{p'}}$. Hence the partial decay width $\Gamma_{\psi p}$ reads
\begin{equation}
\Gamma_{\psi p} = \mathcal{B}_{\psi p}\, \Gamma_r 
=\frac{\bar p_f}{32 \pi^2 M^2_{r}}\frac{1}{2J_{r}+1}  \sum_{\lambda_{R}}\int d\Omega \, | \langle 
\lambda_{\psi}\lambda_{p'} \vert T_{dec} |  
\lambda_{R} \rangle |^2  
= \frac{\bar p_f}{8\pi M_r^2} \frac{6g^2}{2J_r + 1} ,
\label{ampwidth}
\end{equation} 
where  $\mathcal{B}_{\psi p}$ is the $P_c(4450)\to J/\psi p$ branching ratio and $\bar{p}_f$ the $p_f$ momentum at the resonance peak.
Since there are no data in the resonance peak region we can only determine an upper limit 
for $\mathcal{B}_{\psi p}$ within a certain confidence level (CL) to ensure compatibility with the available data.

\section{Results}
\begin{figure}[h]
\begin{minipage}{18pc}
\includegraphics[width=18pc]{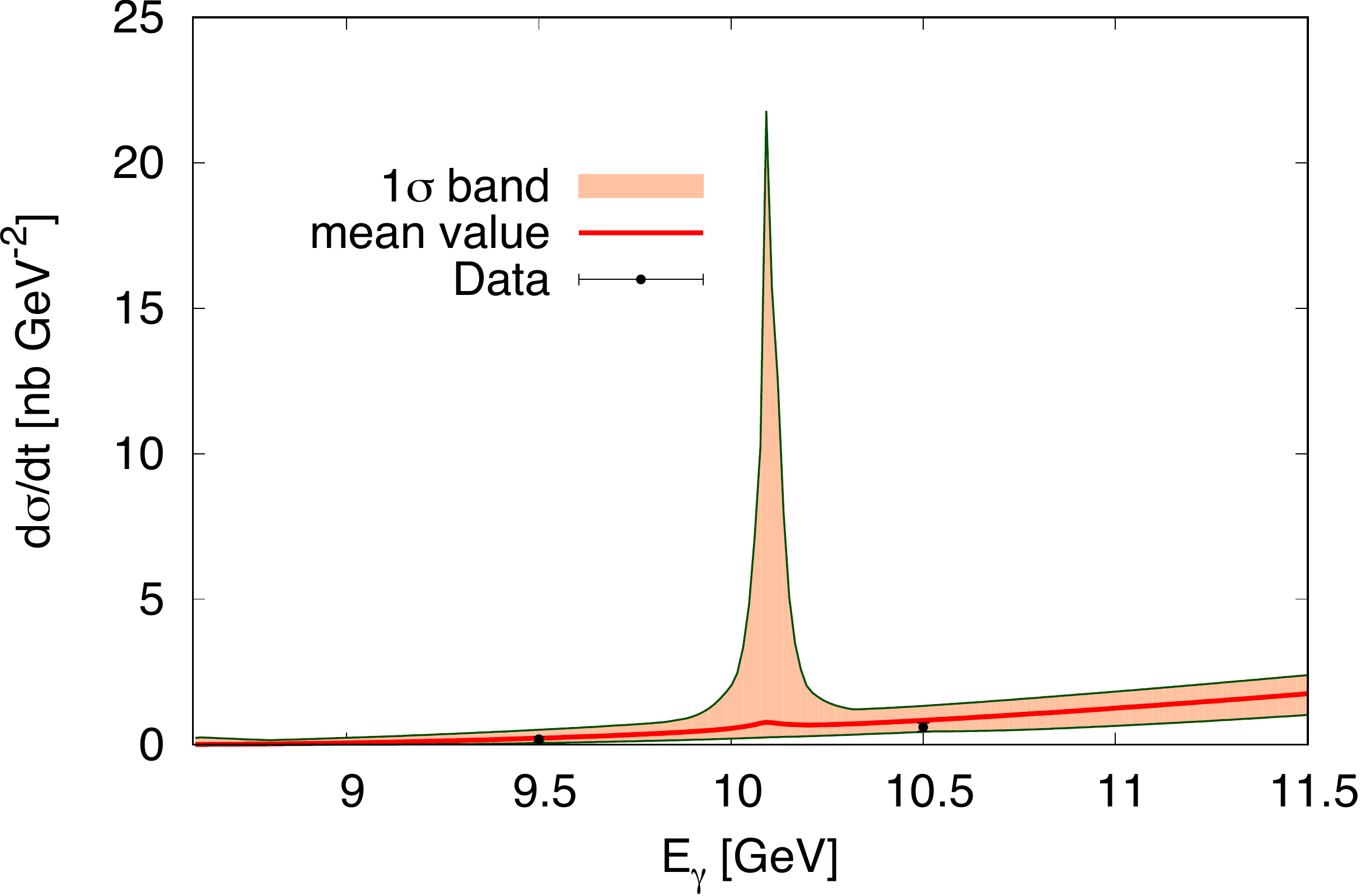}
\end{minipage}\hspace{2pc}%
\begin{minipage}{18pc}
\includegraphics[width=18pc]{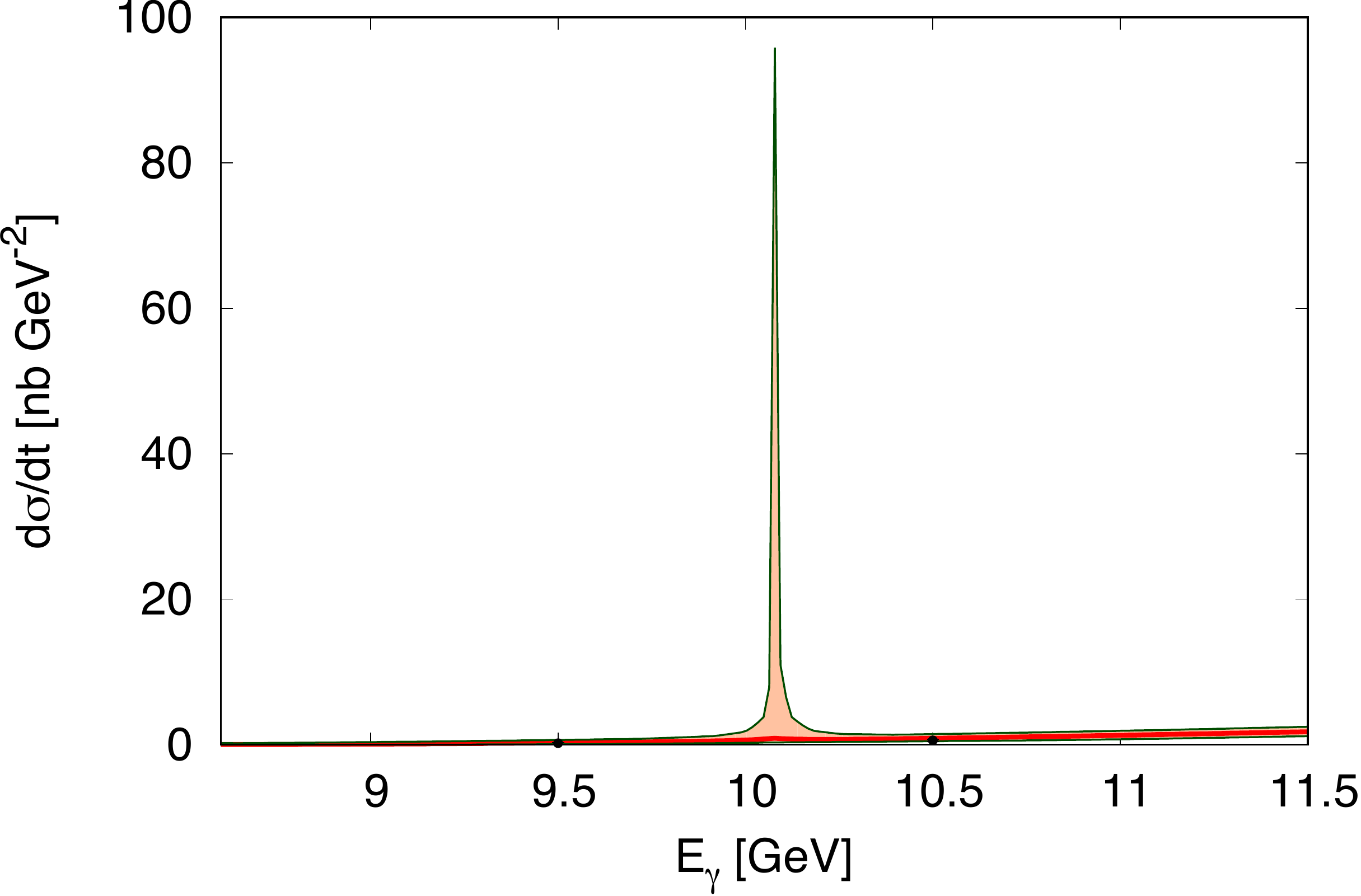}
\end{minipage} 
\caption{Differential cross section in the $P_c(4450)$ peak region at $t=t_\text{min}$
for the  $J_r=3/2^-$ (left plot) and the  $J_r=5/2^+$ (right plot)
spin-parity assignments.} \label{fig:dxs} 
\end{figure}

We fit the available data on the $\gamma p \to J/\psi\, p$ reaction for $|t| \leq 1.5~\text{GeV}^2$ \cite{data}
and we compute the uncertainties in the parameters up to 68\% and 95\% CL using the bootstrap technique \cite{bootstrap}.
We also incorporate the uncertainty of the pentaquark mass and width through a Monte Carlo.
In this way we can propagate exactly the experimental uncertainties to our predictions.
Figure~\ref{fig:dxs} shows the differential cross section $d\sigma \slash dt$ at $t=t_\text{min}$ in the pentaquark energy region for the
$J_r^P=3/2^-$ and $5/2^+$ spin-parity assignments.
The pentaquark peak is very pronounced, which implies that, 
if the LHCb signal is a true resonance and does not couple too weakly to the photon 
(as expected from vector meson dominance), it will be visible in the proposed photoproduction experiment at Jefferson Lab~\cite{PcJLab}.
We found that data allow for the existence of a pentaquark whose $J/\psi p$ 
branching ratio has to be below 30\% for $J^P=3/2^-$ 
and below 17\% for $J^P=5/2^+$ at a 95\% CL. 
The values of the parameters, their uncertainties, 
and the full extent of the analysis can be found in~\cite{ANHB2016a}.

%

\section{Final remarks}
The signals found in the $\Lambda_b^0 \to J/\psi K^- p$ decay by the LHCb collaboration~\cite{LHCbpentaquark}
indicate the possible existence of a hidden charm pentaquark with a mass of $4450~\text{MeV}$.
One of the most promising ways to independently confirm the pentaquark nature of the signal
consists on trying to photoproduce the state using the $\gamma p \to J/\psi p$ reaction.
The recent approval by Jefferson Lab's PAC44 of this very experiment at Hall C~\cite{PcJLab}
and the development of photoproduction models to analize the data~\cite{Wang:2015jsa,ANHB2016a}
make us believe that in the near future we will get a clear answer. 
The codes for the model in~\cite{ANHB2016a} can be downloaded and run online 
from the Joint Physics Analysis Center (JPAC) web page~\cite{JPACWebpageIndiana}.
If the $P_c(4450)$ state is also detected in $J/\psi$ photoproduction,
it would  likely be a compact state.
Then, new experiments aimed to hunt the other members of this hidden-charm pentaquark multiplet can begin~\cite{Bijker17}
as well as experiments to map the $Q^2$ dependence of the pentaquark
electromagnetic form factor and provide insight into its internal structure.

\ack
This work is part of the efforts of the Joint Physics Analysis Center (JPAC) collaboration.
CF-R work is supported in part by 
research grant IA101717 from PAPIIT-DGAPA (UNAM),
by CONACYT (Mexico) grant No. 251817, 
and by Red Tem\'atica CONACYT de F\'{\i}sica en Altas Energ\'{\i}as (Red FAE, Mexico).
ANHB research is supported in part by the Spanish Ministerio de Econom\'{\i}a y Competitividad (MINECO), 
European FEDER funds under contract No. FIS2014-51948-C2-2-P, SEV-2014-0398, 
and Generalitat Valenciana under Contract PROMETEOII/2014/0068.
AP research is supported by the U.S. Department of Energy, Office of Science, 
Office of Nuclear Physics under contract DE-AC05-06OR23177. 
CF-R thanks the organizers for their invitation to the conference and their warm hospitality at Cocoyoc.


\end{document}